\documentclass[aps, prd,twocolumn,showpacs]{revtex4}
\usepackage{graphicx}
\usepackage{latexsym}
\usepackage{amsmath}
\usepackage{psfrag}
\usepackage{amssymb}
\usepackage{color}

\newcommand{\bq}{\begin{equation}}
\newcommand{\eq}{\end{equation}}
\newcommand{\bqn}{\begin{eqnarray}}
\newcommand{\eqn}{\end{eqnarray}}

\begin{document}
\title{Kinetic k-essence ghost dark energy model}

\author{Alberto Rozas-Fern{\'a}ndez$^1$}
\email{Alberto.Rozas@port.ac.uk}
\affiliation{Institute of Cosmology and Gravitation, University of Portsmouth, Portsmouth, PO1 3FX, UK}
\date{\today}
\begin{abstract}

A ghost dark energy model has been recently put forward to explain the
current accelerated expansion of the Universe. In this model, the  energy
density of ghost dark energy, which comes from the Veneziano ghost
of QCD, is proportional to the Hubble parameter, $\rho_D=\alpha H$. Here $\alpha$ is a constant of order $\Lambda^3_{QCD}$ where $\Lambda_{\rm QCD}\sim 100 MeV$ is the QCD
mass scale. We consider a connection between ghost dark energy with/without interaction between the components of the dark sector and the kinetic k-essence field. It is shown that the cosmological evolution of the ghost dark
energy dominated Universe can be completely described a kinetic k-essence scalar field. We reconstruct the kinetic k-essence function $F(X)$ in
a flat Friedmann-Robertson-Walker Universe according to the evolution of
ghost dark energy density.

\end{abstract}

 \maketitle

\section{Introduction}

Nowadays, there is little doubt that the Universe is currently undergoing an acceleration of its expansion. This is supported by the overwhelming evidence provided by cosmological observations from Type Ia supernovae (SN Ia)
\cite{SN}, Cosmic Microwave Background (CMB) anisotropies measured
with the WMAP satellite \cite{CMB}, Large Scale Structure
\cite{LSS}, weak lensing \cite{WL} and the integrated Sach-Wolfe
effect \cite{ISWE}. In order to explain why this happens, it is usual to postulate the existence of a substance, dubbed dark energy (DE), that behaves as if it had negative pressure and is responsible for this present cosmic acceleration. See \cite{nederev, Cai:2009zp} for a recent reviews on DE models.
Nevertheless, the underlying physical
mechanism behind this phenomenon remains unknown what has led to explore other possibilities such us the quantum cosmic model \cite{benigner} or $f(R)$
theories \cite{Nojiri:2010wj}.

A new model of DE, under the name of Veneziano ghost dark energy (GDE), has been recently proposed \cite{Urban,Ohta}. Veneziano ghost is supposed to exist for solving the $U(1)_{A}$ problem in low-energy effective theory of QCD \cite{Witten,Veneziano,RST,NA,KO}, although it is completely decoupled from the physical sector \cite{Ohta2}. The central point in the GDE model is that the Veneziano ghost, being unphysical in the QFT formulation in the Minkowski spacetime, exhibits important non-trivial physical effects in an expanding Universe such as our Friedmann-Robertson-Walker (FRW) Universe, or in a spacetime with non-trivial topological structure. These effects are naturally small and give rise to a vacuum energy density $\rho_{D} \sim \Lambda_{QCD}^{3}H \sim (10^{-3} {\text eV})^{4}$, where $\Lambda_{QCD}$ is the QCD mass scale and $H$ is the Hubble parameter \cite{Ohta}. With $\Lambda_{QCD}\sim100MeV$ and $H\sim10^{-33}eV$, this small vacuum energy density has precisely the right value to be the driving force accelerating the Universe today \cite{Urban,Ohta,CaiGhost}. This remarkable coincidence implies that the GDE model is free from the fine tuning problem \cite{Urban,Ohta}. In addition, the appearance of the QCD scale could be relevant for a solution to the cosmic coincidence problem, as it may be the scale at which dark matter (DM) forms~\cite{forbes}. In general, it is very difficult to accept such a linear behaviour in the energy density because QCD is a theory with a mass gap determined by the  scale $\sim100MeV$. Therefore, it is generally expected that there should be exponentially small corrections rather than linear corrections $\sim H$. In Refs. \cite{Zhitnitsky:2010ji,Holdom:2010ak, Zhitnitsky:2011tr} this question has been elaborated in detail where it has been argued that the linear scaling $\sim H$ is due to the complicated topological structure of strongly coupled QCD, not related to the physical massive propagating degrees of freedom. The advantage of the GDE model, when compared to other DE models, is that the DE can be totally explained within the standard model and general relativity, without resorting to any new field, new degree(s) of freedom, new symmetries  or modifications of gravity \cite{CaiGhost}. The thermodynamics of the GDE model has been studied in \cite{Feng:2011ev}
. Furthermore, the GDE model has been fitted with current observational data including SNIa, BAO,
CMB, BBN and Hubble parameter and although it was found that the current data do not favour the GDE model when compared to the
$\Lambda$CDM model \cite{CaiGhost}, the result is not conclusive and further study is needed. We also note that models with similar terms in the Friedmann equation were introduced some time ago in frames of inhomogeneous or imperfect fluids (see \cite{Nojiri:2005sr,Nojiri:2006ri,Capozziello:2005pa}).

One clarification is in order since there are some dark energy models where the ghost plays the role of dark energy (see, e.g., \cite{Piazza:2004df}) and becomes a real propagating physical degree of freedom subject to some severe constraints. However, the Veneziano ghost is not a new physical propagating degree of freedom and therefore the GDE model does not violate unitarity, causality, gauge invariance and other important features of renormalizable quantum field theory, as advocated in \cite{Zhitnitsky:2010ji,Holdom:2010ak, Zhitnitsky:2011tr}. At the same time, it has been emphasized in \cite{Urban,Ohta}, and especially later in \cite{Zhitnitsky:2010ji,Holdom:2010ak, Zhitnitsky:2011tr}, that in fact the description of GDE in terms of the Veneziano ghost is just a matter of convenience to describe very  complicated infrared dynamics of strongly coupled QCD.  One can describe the same dynamics using some other approaches (e.g. direct lattice simulations) without using the ghost.

On the other hand, as is well known, scalar field models are an effective
description of an underlying theory of DE. From a given equation of state (EoS) describing a fluid, a scalar field model can be derived (see e.g. \cite{Ellis:1990wsa}). This correspondence, however, is not in general one to one: scalar fields have an extra degree of freedom and therefore show a dynamics that is more involved. The question is to what extent the model admits a solution that acts as an attractor, so that it can mimic the fluid evolution thereby avoiding a strong fine tuning problem that would render the model useless. Scalar fields are popular
not only because of their mathematical simplicity
and phenomenological richness, but also because they
naturally arise in particle physics including supersymmetric field
theories and string/M theory. However, these fundamental theories
do not predict their potential $V(\phi)$ or kinetic term uniquely. Our aim is to investigate whether a minimally coupled scalar field with a specific Lagrangian can mimic the dynamics of the GDE model so that this model can be related to some fundamental theory, as it is for a scalar field. For this task, it is then meaningful to reconstruct the $V(\phi)$ or the kinetic term of a DE model possessing some significant features of the underlying theory of DE, such as the GDE model. In order to do that, we establish a correspondence between the scalar field and the GDE by identifying their respective energy densities and equations of state and then reconstruct the potential (if the scalar field is quintessence or the tachyon, for instance) or the kinetic term (k-essence belongs to this class) and the dynamics of the field. Some work has already been done in this
direction. Quintessence and tachyonic GDE models
have been discussed in \cite{Sheykhi:2011nb} and \cite{Sheykhi:2011fe},
respectively. In this paper, within the different candidates to play the role of the DE, we have chosen the kinetic k-essence, as this has emerged
as a possible source of DE \cite{AMS1, AMS2} where the cosmic acceleration can be realised by the kinetic energy $X$ of the field $\phi$. For instance, the correspondence between the kinetic k-essence field
and the holographic DE model
was already explored in \cite{Cruz:2008cwa}. Quintessence and tachyonic models belong to k-essence. In k-essence, the higher order
terms are not necessarily negligible which, interestingly, can give rise to new dynamics not possible in quintessence. Every quintessence model can be viewed as a k-essence model generated by a kinetic linear function. On the other hand, the tachyon model is classified as k-essence because it belongs to a class of the action for the k-essence. However, in the sense that the kinetic energy of the tachyon needs to be suppressed to realise cosmic acceleration, this scenario is different from k-essence.  Finally, an advantage
of purely kinetic k-essence Lagrangians, thanks to its technical naturalness as from a shift symmetry, is that
possess a single degree of freedom, $\mathcal{L}= F(X)$, like quintessence.

The rest of the paper can be outlined as follows. In the next section we reconstruct the kinetic k-essence GDE model
in the light of the GDE. In section III we extend the study to the
interacting GDE model by using the latest data from observations. The
conclusions are drawn in Sec. IV.

\section{Kinetic k-essence ghost dark energy model }
Let us start with a spatially flat FRW Universe filled with a matter component and GDE. The
Friedmann equation, which governs its dynamics, reads

\begin{eqnarray}\label{Fried}
H^2=\frac{1}{3M_P^2} \left( \rho_m+\rho_D \right),
\end{eqnarray}
where $\rho_m$ is the energy density of pressureless DM and $\rho_D$ is the GDE density.

By introducing the
dimensionless density parameters
\begin{equation}\label{Omega}
\Omega_m=\frac{\rho_m}{\rho_{cr}},\ \ \
\Omega_D=\frac{\rho_D}{\rho_{cr}},\ \
\end{equation}
where $\rho_{cr}={3M_P^{2}H^{2}}$ is the critical energy density,
the Friedmann equation can also be written as
\begin{equation}\label{fridomega}
\Omega_m+\Omega_D=1.
\end{equation}
Equivalently, Eq. (\ref{Fried}) can be expressed as
\begin{eqnarray}  \label{Fridmann1}
H(z)= H_{0}\left [ \frac{\Omega_{m,0}(1+z)^{3}}{1-\Omega_{D}}
\right ]^{1/2}
\end{eqnarray}
where $z=(1/a)-1$ is the redshift and $H_{0}$ and $\Omega_{m,0}$ are the current values for $H$ and
$\Omega_{m}$.

The GDE density is
proportional to the Hubble parameter \cite{Ohta}
\begin{equation}\label{GDE}
\rho_D=\alpha H.
\end{equation}
where $\alpha$ is a constant of order $\Lambda_{\rm QCD}^3$ and
$\Lambda_{\rm QCD}\sim 100 MeV$ is the QCD mass scale.

The conservation equations are given by
\begin{eqnarray}
\dot\rho_m+3H\rho_m&=&0,\label{consm}\\
\dot\rho_D+3H\rho_D(1+w_D)&=&0\label{consd}.
\end{eqnarray}
If we now take the time derivative of the GDE density (\ref{GDE}) and use the
Friedmann equation we arrive at
\begin{equation}\label{dotrho1}
\dot{\rho}_D=\rho_D \frac{\dot{H}}{H}=-\frac{\alpha }{2 M_P^2}
\rho_D(1+r+w_D)
\end{equation}
where
\begin{equation}\label{u1}
r=\frac{\rho_m}{\rho_D}=\frac{\Omega_m}{\Omega_D}=\frac{1-\Omega_D}{\Omega_D}
\end{equation}
is the ratio of the energy densities. By substituting the relation (\ref{dotrho1}) into the
continuity equation (\ref{consd}), and after using (\ref{u1}) we get the dynamical equation of state (EoS) of GDE
\begin{equation}\label{wD1}
w_D=-\frac{1}{2-\Omega_D}.
\end{equation}

\begin{figure}[htp]
\psfrag{z}{$z$} \psfrag{wD}{$w_{D}$}
\includegraphics[scale=1.02]{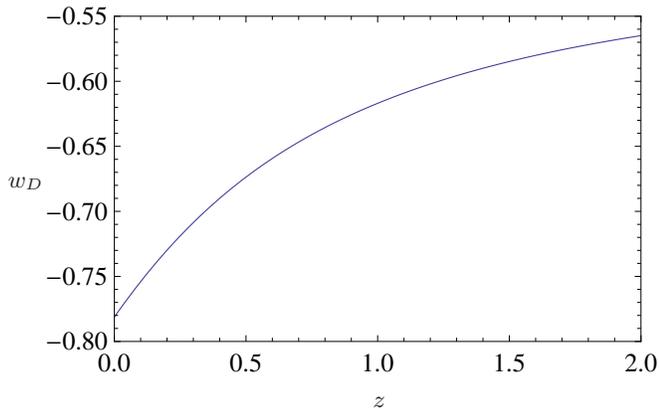}
\caption{The evolution of the equation of state, $w_{D}$, for the ghost dark energy model.}
\end{figure} At early times, when the DE is negligible, $\Omega_D\ll 1$, we have $w_D=-1/2$. On the other hand, at late times, DE dominates, $\Omega_D\rightarrow 1$, and the GDE behaves as a
cosmological constant with $w_D= -1$ entering
the de Sitter phase in the far future. The SN Ia observations have provided information of the cosmic expansion history around the redshift $z\leq2$ by the measurement of luminosity distances of the sources, which is also the range spanned by direct measurements of the Hubble parameter. Therefore, we plot all the figures in this paper in the redshift range between $z=0$ and $z=2$. In Fig. 1 the
evolution of $w_D$ is shown as a function of  $z$. It is clear to see that $w_D$ evolves in the region $w\geq-1$, so the GDE model cannot realise the
phantom crossing. It is important to mention that in this model, unlike other models of DE, the evolution of $w_D$ is completely determined by the dynamics of $\Omega_D$. We notice that $w(z=0)=w_{0}\approx-0.78$, which is $2\sigma$ off according to WMAP ~\cite{wmap}.  However, our equation of state is dynamical so this statement is significantly weakened in this case.  Indeed, WMAP in this case only measures the integrated equation of state from last scattering to now, which does not provide much information on its effective time dependence~\cite{bruce}.  Furthermore, the usual ansatz for the linear redshift dependence, $w(z) = w_0 + w_a (1-a)$, does not necessarily provide a good fit for the behaviour shown in Fig.1.

\begin{figure}[htp]
\psfrag{z}{$z$}
\includegraphics[scale=1.02]{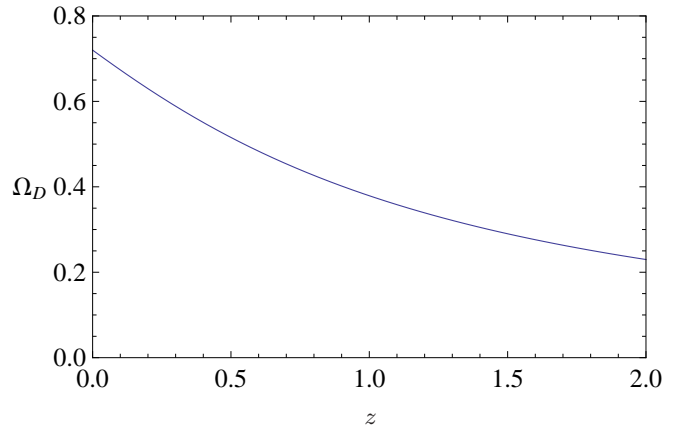}
\caption{The evolution of $\Omega_{D}$ for the ghost dark energy model.}
\end{figure}

If we now take the time derivative of $\Omega_D=\alpha/(3M_P^2H)$, we
get ${\dot{\Omega}_D}= -H(1+z) \frac{d\Omega_D}{dz}$ which allows us to
get
\begin{equation}\label{Omegaprime1}
\frac{d\Omega_D}{dz}=\frac{\Omega_D}{(1+z)}\frac{\dot H}{H^2}.
\end{equation}
Combining the Friedmann equation with Eqs. (\ref{u1}) and (\ref{wD1})
we obtain the differential equation that governs the whole dynamics of the GDE \cite{shmov}
\begin{equation}\label{Omegaprime2}
\frac{d\Omega_D}{dz}=-\frac{3}{(1+z)}\Omega_D\left(
\frac{1-\Omega_D}{2-\Omega_D}\right).
\end{equation}

In Fig. 2 we have plotted the evolution of $\Omega_{D}$ as a function of $z$, where $\Omega_{D,0}=0.72$ is the value of $\Omega_{D}$ at the present epoch. It can be seen that DE dominates at late times.

\begin{figure}[htp]
\psfrag{z}{$z$} \psfrag{q}{$q$}
\includegraphics[scale=1.02]{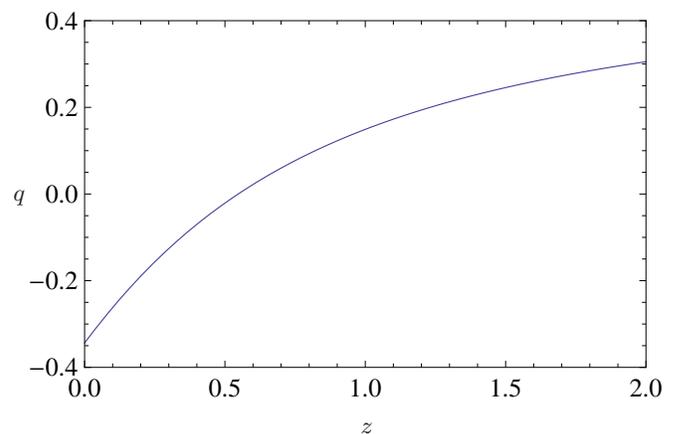}
\caption{The evolution of the deceleration parameter, $q$, for the ghost dark energy model.}
\end{figure}

We can also compute the deceleration parameter

\begin{equation}\label{q1}
q=-1-\frac{\dot{H}}{H^2}=\frac{1}{2}-\frac{3}{2}\frac{\Omega_D}{(2-\Omega_D)}.
\end{equation} where Eqs. (\ref{Omegaprime1}) and (\ref{Omegaprime2}) have been used after the second equal sign. The deceleration parameter $q$ develops from $\frac{1}{2}$ at early times, when $\Omega_D\ll 1$, to $-1$ at late times when DE dominates and $\Omega_D\rightarrow 1$. The behaviour of $q$ is plotted in Fig. 3 where the transition from deceleration to acceleration is seen to occur at $z\simeq
0.56$. Also, when we take $\Omega_{D,0}=0.72$, the value of the deceleration parameter at the present era is $q \approx-0.34$, well in agreement with recent observational data \cite{daly}.

We are now ready to establish the correspondence between GDE and the kinetic k-essence scalar field.
As a DE candidate, k-essence \cite{COY, AMS1, AMS2} is usually defined as a scalar field $\phi$ with a
non-canonical kinetic energy associated with a Lagrangian
$\mathcal{L} = -V(\phi) F(X)$, where $V$ is the potential and $X=\frac{1}{2}\partial_{\mu}
\phi\partial^{\mu} \phi$ is the kinetic term. In the case of the k-essence scalar field, the negative pressure that explains the accelerated expansion
arises out of modifications to the kinetic energy.

K-essence models are described by an effective
minimally coupled scalar field with a non-canonical term. If for a
moment we neglect the part of the Lagrangian containing ordinary
matter, the general action for a k-essence field $\phi$ minimally
coupled to gravity is

\begin{equation}\label{actionKessence}
S= S_{G}+S_{\phi}=- \int d^{4}x \sqrt{-g}\left(\frac{R}{2}+
F_k(\phi,X)\right),
\end{equation}
where $F_k(\phi,X)$ is an arbitrary function of $\phi$ and its kinetic energy $X$. A possible motivation for actions of this form comes from considering low-energy effective string
theory in the presence of a high-order derivative terms.

In what follows, we
shall work with the simple k-essence models for which
the potential $ V=V_{0}=$ constant \cite{Scherrer:2004au}. We also assume that $V_{0}=1$
without any loss of generality. One reason for studying k-essence
it that it is possible to construct a particularly interesting
class of such models in which the k-essence energy density tracks
the radiation energy density during the radiation-dominated era,
but then evolves toward a constant-density DE component
during the matter-dominated era. Such behaviour can to a certain
degree solve the coincidence problem \cite{COY, AMS1, AMS2}. Because of this dynamical attractor behaviour, the cosmic evolution is insensitive to initial conditions. Another feature of k-essence is that it changes its speed of evolution in dynamic response to changes in the background equation of state.

We now restrict ourselves to the subclass of kinetic k-essence,
with an action independent of $\phi$
\begin{eqnarray}\label{actionKessence}
S= -\int d^{4}x \sqrt{-g} F(X).
\end{eqnarray}

The consideration of constraints on purely kinetic k-essence models from the latest observational data by applying model comparison statistics (F-test, $AIC_{c}$, and BIC) has found that these models are favoured over the $\Lambda$CDM by the combined data \cite{Yang:2009zzl}.

We assume a FRW metric $ds^2 = dt^2 -
a^2(t)\, d\vec{x}^2$ and consider
$\phi$ to be smooth on scales of interest so that $X = \frac{1}{2}
\dot{\phi}^2\geq0$. The energy-momentum tensor of the k-essence is
obtained by varying the action (\ref{actionKessence}) with respect
to the metric, yielding
\begin{eqnarray}\label{energy-momentKessence}
T_{\mu\nu}=  F_{X}
\partial_{\mu}\phi\partial^{\mu}\phi - g_{\mu\nu}F,
\end{eqnarray}
where the subscript $X$ denotes differentiation with respect to
$X$. Identifying (\ref{energy-momentKessence}) with the
energy-momentum tensor of a perfect fluid we have the k-essence
energy density $\rho_{\phi}$ and pressure $p_{\phi}$

\begin{eqnarray}\label{RoKessence}
\rho_{\phi}= F-2XF_{X}
\end{eqnarray}
and
\begin{eqnarray}\label{PressureKessence}
p_{\phi}=-F.
\end{eqnarray}

Assuming as usual that the energy density is
positive, we have that $F-2 X F_X> 0$.
The equation of state for the
k-essence fluid can be written as $p_{\phi} = w_{\phi}\rho_{\phi}$
with with $F>0$,
\begin{eqnarray}\label{wKessence}
w_{\phi} = \frac{p_{\phi}}{\rho_{\phi}}= \frac{F}{2XF_{X}-F}.
\end{eqnarray} As long as the condition $|2XF_{X}|\ll|F|$ is satisfied, $w_{\phi}$ can be close to $-1$.

Besides, the effective sound speed for the kinetic k-essence, which is the quantity relevant for the growth of perturbations, is expressed as
\begin{eqnarray}\label{sspeed}
c^{2}_{\rm s}=\frac{\partial p_{\phi}/\partial
X}{\partial\rho_{\phi} /\partial
X}=\frac{F_{X}}{F_{X}+2XF_{XX}}=\frac{F_{X}^{2}}{(XF_{X}^{2})_{X}},
\end{eqnarray}
where $F_{XX}\equiv d^{2}F/dX^{2}$. The definition of the sound
speed comes from the equation describing the evolution of linear
adiabatic perturbations in a k-essence dominated Universe
\cite{Garriga:1999vw} (the non-adiabatic perturbation was discussed
in \cite{Unnikrishnan:2010ag}, here we only consider the case of
adiabatic perturbations). Perturbations can become unstable if the
sound speed is imaginary, $c_{\rm s}^2<0$ and there would be unpleasant consequences for structure formation. This is not, however, the case for the Veneziano ghost since it is not a physical propagating degree of freedom and therefore the notion of the speed of sound does not exist in this context. A potentially interesting requirement to consider is $c_{\rm s}^2
\leq 1$, which says that the sound speed should not exceed the
speed of light, which suggests violation of causality. Though this
is an open problem (see e. g.
\cite{Babichev:2007dw,Bruneton:2006gf,Kang:2007vs,Bonvin:2006vc,Gorini:2007ta,Ellis:2007ic}). It is important to notice that the k-essence models constructed to solve the coincidence problem inevitably give rise to the superluminal propagation of the field ($c_{\rm s}^2>1$) at some stage of the cosmological evolution \cite{Bonvin:2006vc}.

\begin{figure}[htp]
\psfrag{z}{$z$} \psfrag{F}{$F$}
\includegraphics[scale=1.02]{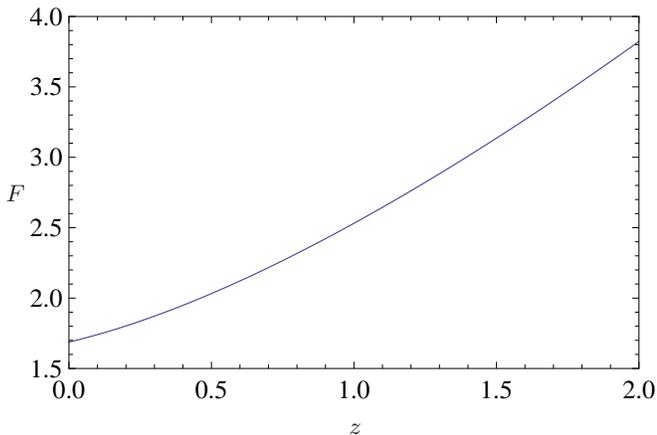}
\caption{Variation of $F(z)$ in units of $\frac{\alpha^{2}}{3M_{P}^{2}}$ for the ghost dark energy model.}
\end{figure}
\begin{figure}[htp]
\psfrag{z}{$z$} \psfrag{X}{$X$} \psfrag{X0}{$X_{0}$}
\includegraphics[scale=1.02]{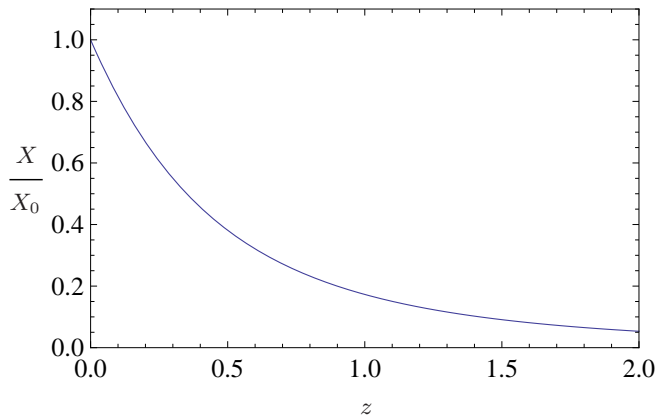}
\caption{Variation of $\frac{X}{X_{0}}(z)$ in units of $M_{P}^{2}H_{0}^{2}$ for the ghost dark energy model.}
\end{figure}

For a flat FRW metric, applying the Euler-Lagrange equation for
the field to the action (\ref{actionKessence}) we find the
equation of motion for k-essence field
\begin{eqnarray}\label{phiequation}
(F_{X}+2XF_{XX}) \ddot{\phi}+ 3HF_{X}\dot{\phi}=0,
\end{eqnarray}
which can be rewritten in terms of $X$ as
\begin{eqnarray}\label{phiequation1}
(F_{X}+2XF_{XX}) \dot{X}+ 6HF_{X}X=0.
\end{eqnarray}
 If we now change
the independent variable from time $t$ to the scale factor $a$, we
obtain
\begin{eqnarray}\label{phiequation2}
(F_{X}+2XF_{XX})\,a\,\frac{dX}{da}+ 6F_{X}X=0.
\end{eqnarray}
This equation can be integrated exactly, for arbitrary $F$,
yielding
\begin{eqnarray}\label{phiequation3}
XF_{X}^{2}=ka^{-6}=k(1+z)^{6},
\end{eqnarray}
where $k$ is a constant of integration \cite{Scherrer:2004au}. This solution had been previously derived in a different form in \cite{Chimento:2003ta}. Given a function
$F(X)$, Eq.(\ref{phiequation3}) allows us to find solutions $X(z)$
and then the other parameters of the k-essence fluid like
$\rho_{\phi}$, $p_{\phi}$, $w_{\phi}$ and $c_{\rm
s}^2$ as a function of the
redshift, $z$. The stable nodes of Eq.~(\ref{phiequation2}) were analysed in \cite{Bertacca:2007ux}, corresponding to solutions for which either $\partial F/\partial X|_{X_{*}}=0$ or $X_{*}=0$ (both with $w_{\phi}=-1$).

From Eqs.(\ref{RoKessence}),(\ref{wKessence}) and (\ref{Fried}), we can obtain the expression for $F$ as a function of the $z$
\begin{eqnarray}\label{Fdez}
F(z)= -\rho_{\phi}\,w_{\phi}=-
3M_{p}^{2}\,H^{2}(z)\,\Omega_{\phi}(z)\,w_{\phi}(z).
\end{eqnarray} which is positive since $w_{\phi}(z)<0$. As we have demanded that
the energy density be positive, Eq.(\ref{RoKessence}) implies that
$F_{X}<F/2X$. Therefore, for kinetic k-essence, $F>0$ and $F_X <
0$ imply that $w_{\phi} > -1$.

At this point, we focus on the reconstruction of $F(X)$ in the
redshift range between redshift $z=0$ and $z=2$ for contrasting the model against the data.
\begin{figure}[htp]
\psfrag{z}{$z$} \psfrag{F}{$F$} \psfrag{X}{$X$} \psfrag{X0}{$X_{0}$}
\includegraphics[scale=1.02]{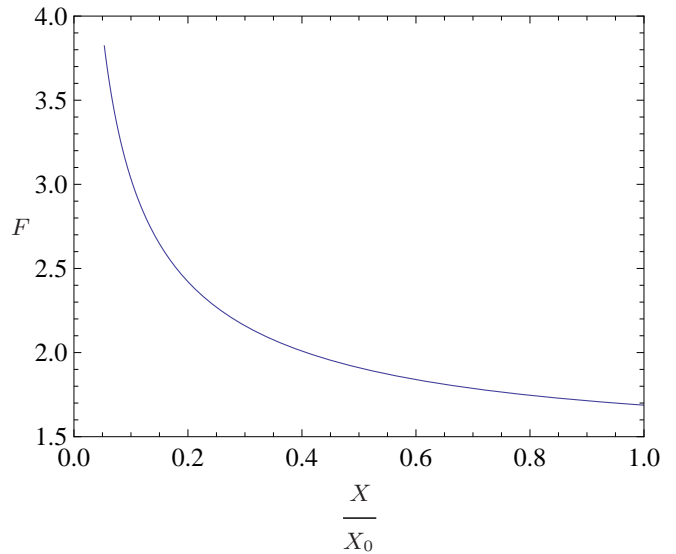}
\caption{Reconstructed $F(X/X_{0})$ for the ghost dark energy model.}
\end{figure}
In order to establish the correspondence between the GDE and
the kinetic k-essence, we identify $\rho_{\phi}$
with $\rho_{D}$ and $w_D$ with $w_\phi$.

By making use of Eqs. (\ref{GDE}), (\ref{wD1}) and (\ref{Omega}), we can then express Eq. (\ref{Fdez}) as
\begin{eqnarray}\label{Fvsz}
F(z)= \frac{\alpha^{2}}{3M_{P}^{2}\Omega_{D}(2-\Omega_{D})}.
\end{eqnarray}
Also, Eqs. (\ref{GDE}), (\ref{wD1}),(\ref{Omega}) and (\ref{Fridmann1}) provide an expression for $X$ as a function of $z$
\begin{equation}\label{XdezexplicitX0anal}
\frac{X}{X_{0}}(z)=\left[\frac{\Omega_{D}(\Omega_{D,0}-2)}{\Omega_{D,0}(2-\Omega_{D})}\right]^{2}
\end{equation}
where $X_{0}$ and $\Omega_{D,0}$ are the current values for $X$ and
$\Omega_{D}$, respectively. The stable node of Eq.~(\ref{phiequation2}) is the one for which $X_{0}=0$ and $w_{\phi}=-1$.

Finally, from Eqs. (\ref{Fdez}) and (\ref{XdezexplicitX0anal}) we obtain the function $F=F(X/X_{0})$

\begin{eqnarray}\label{FdeXexplicit}
F(X/X_{0})= \frac{\alpha^{2}\Omega_{D,0}}{3M_{P}^{2}(2-\Omega_{D,0})\Omega_{D}^{2}}\sqrt{\frac{X}{X_{0}}}.
\end{eqnarray}

The behaviour of $F$ as a function of $z$ is shown in Fig. 4 where we can see that $F$ is positive for an accelerating
Universe with GDE (as it must necessarily be from Eq. (\ref{Fdez})). Likewise, the evolution of $X/X_{0}$ as a function of $z$ is shown in Fig. 5.

The kinetic k-essence GDE, represented by the function
$F$, is plotted in Fig. 6 as a function of $X/X_{0}$.
From Figs. 5 and 6 we can see the dynamics of the kinetic k-essence
field explicitly. From Fig. 6 we can see that the reconstructed $F=F(X/X_{0})$ is a monotonically  decreasing function of
$X/X_{0}$, well-behaved and a single
valued function in the relevant redshift range. This is because for $X>0$, the
sign of $\frac{F_{X}}{F}$ is related to the value of $w_{\phi}$.
It is important to indicate that the reconstruction of $F(X/X_{0})$ only
involves the portion of it over which the field evolves to give
the required $H(z)$. The behaviour of this model approaches a de Sitter phase in
the late time.

\section{Interacting kinetic k-essence ghost dark energy model}

To further study the GDE model, we shall consider a possible interaction between DM and DE. The interacting models were first proposed by Wetterich in an attempt to lower down the value of the cosmological term \cite{Wetterich:1987fm}. Later on, it was proved to be efficient in alleviating the cosmic coincidence problem \cite{Amendola:1999er, coincidence} unlike the $\Lambda$CDM which cannot address it. In addition, the interaction may not only be likely but inevitable \cite{Brax:2006kg}. On the other hand, ignoring the interaction may result in a misled interpretation of the data regarding the equation of state of DE. In fact, it has been shown that a measured phantom equation of state may be mimicked by an interaction \cite{Das:2005yj, Amendola:2006ku}.

\begin{figure}[htp]
\psfrag{z}{$z$} \psfrag{wD}{$w_{D}$} \psfrag{b2}{$b^{2}$}
\includegraphics[scale=0.855]{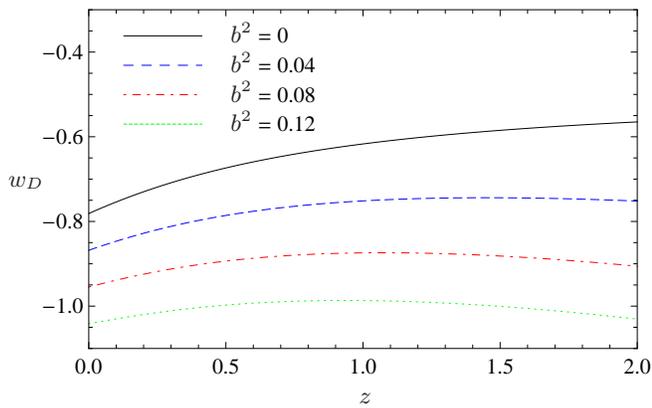}
\caption{The evolution of the equation of state, $w_{D}$, for the interacting ghost dark energy model.}
\end{figure}

It is usually assumed that both DM and DE only couple
gravitationally. However, given their unknown nature and that the
underlying symmetry that would set the interaction to zero is
still to be discovered, an entirely independent behaviour between
them would be very special indeed. Assuming that the DE is a field, it would be more natural for it to couple with the remaining fields of the theory, in particular with DM, as it is quite a general fact that different fields generally couple. Moreover, since DE
gravitates, it must be accreted by massive compact objects such as
black holes and, in a cosmological context, the energy transfer
from DE to DM may be small but non-vanishing.

\begin{figure}[htp]
\psfrag{z}{$z$} \psfrag{b2}{$b^{2}$}
\includegraphics[scale=0.90]{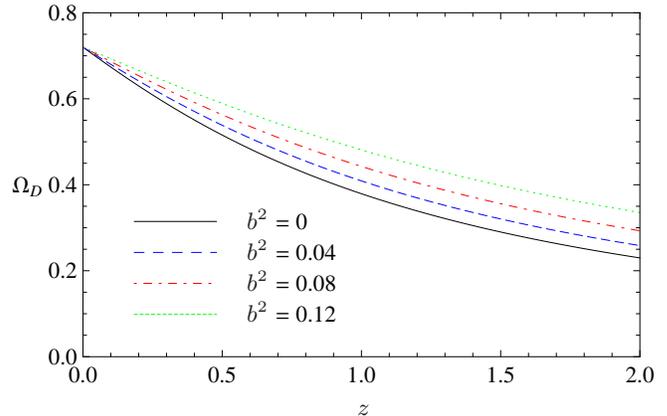}
\caption{The evolution of the equation of state, $\Omega_{D}$, for the interacting ghost dark energy model.}
\end{figure}
In spite of the fact that the available empirical data cannot
discriminate between the existence of a small interaction and its total absence, there are some indications that favour the interaction: (i) The interaction affects the time required for a self-gravitating, collapsing, structure to reach equilibrium as well as the equilibrium
configuration itself, which implies that the Layzer-Irvine equation \cite{Peebles:1994xt, Layzer} has to be generalised to include the interaction. In this regard, it has been
observed a small interaction when studying the dynamics of a group of galaxy clusters \cite{Abdalla:2007rd, Bertolami:2007zm}. (ii) Since the interaction modifies the rate of evolution of the metric potentials, the integrated Sachs-Wolfe (ISW) component of the CMB radiation is enhanced. In fact, it has been recently disclosed that the late ISW effect has the unique ability to provide an insight into the coupling \cite{hePRD09}. The cross-correlation of galaxy catalogs with CMB maps also suggests a small interaction \cite{Olivares:2008bx}. A number of studies have been devoted to analyse the constraints on the interaction from the probes of the cosmic expansion history by using the WMAP, SNIa, BAO and SDSS data, etc.
\cite{71, 72, 20, 26, 75, 76, hePRD09, pp}. Complementary probes of the coupling have been carried out in the study of the growth of cosmic structure
\cite{31, Caldera, cb, bb}. It has been also found that a non-zero interaction leaves a clear change in the growth index
\cite{31,Caldera}. See also Ref. \cite{Shafieloo:2010zz}
 for a possible decay of DE into DM consistent with the data.
\begin{figure}[htp]
\psfrag{z}{$z$} \psfrag{b2}{$b^{2}$} \psfrag{q}{$q$}
\includegraphics[scale=0.91]{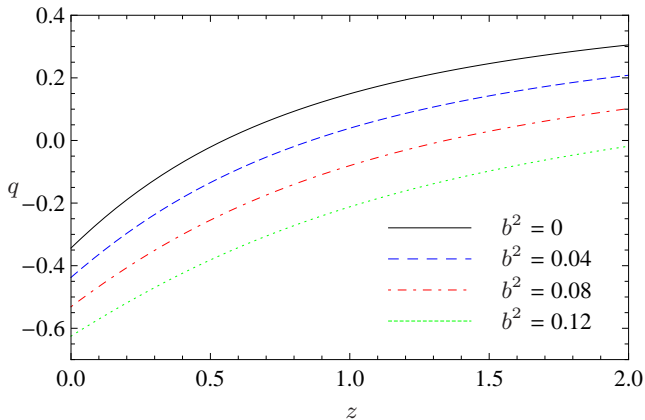}
\caption{The evolution of the deceleration parameter, $q$, for the interacting ghost dark energy model.}
\end{figure}

If DE couples to DM through some interaction, this affects the past expansion history
 of the Universe as well as the cosmic structure formation. The matter
density, $\rho_{m}$, drops more slowly than $a^{-3}$. A slower
matter density evolution fits the supernovae data as well as the
$\Lambda$CDM concordance model does \cite{coincidence}. The
interaction also alters the age of the Universe, the evolution of
matter and radiation perturbations and gives rise to a different
matter and radiation power spectra. In the absence of a fundamental
theory for DE, the coupling term cannot be derived from
microphysics. Most studies on the interaction between dark sectors
rely either on the assumption of interacting fields from the
outset \cite{Amendola:1999er, Das:2005yj}, or from
phenomenological requirements \cite{Zimdahl:2001ar}. The aforesaid
interaction has also been considered from a thermodynamical
perspective \cite{Wang:2007ak, Pavon:2007gt} and has been shown
that the second law of thermodynamics imposes an energy transfer
from DE to DM. Other authors
have analysed the possibility of having DM decaying into
DE but it is required to have at least one of the fluids with a
non null chemical potential, an assumption that which we believe
it is not completely acceptable as it introduces too many
unjustified components, and also relies on the assumption of a
nearly standard evolution of pertubations on interacting DE models \cite{Pereira:2008at}. On the other hand, having a
energy transfer from DM to DE would worsen the
coincidence problem \cite{delCampo:2008jx}.

In the case of an interaction between GDE and DM, their energy
densities no longer satisfy independent conservation laws. They
obey instead
\begin{equation}\label{diff1}
\dot{\rho}_m+3H\rho_m = Q \;,
\end{equation}
\begin{equation}\label{diff2}
 \dot{\rho}_D+3H(1+w_D)\rho_{D} = -Q,
\end{equation}where $Q$ is an interaction term whose form is not unique.

The time derivative of Eq. (\ref{u1}) is $\dot{r}=-\dot{\Omega_{D}}/\Omega_{D}^{2}$. Moreover, from Eqs. (\ref{diff1}) and (\ref{diff2}) we obtain
\begin{equation}\label{rpunto}
\dot{r}=3Hrw_{D}+\frac{Q}{\rho_{D}}(1+r).
\end{equation}
Considering $w_{D}<-{1\over 3}$ and Eq. (\ref{rpunto}) we can see that
in the non-interacting kinetic k-essence GDE model $\dot{r}<-Hr$ and $\ddot{a}>0$ cannot be achieved
simultaneously (see \cite{Hu:2006ar}). Therefore, in contrast to the recent data which indicate
$r\sim \mathcal{O}(1)$,
$r\to 0$ eventually. This can be considered as an important hint for the need of interacting DE.

The expression for $Q$ must be small (at least lower than
$3H\rho_m$) because if it were large and positive, DE
would not dominate the expansion today. On the other hand, if $Q$
were large and negative, the Universe would have been dominated by
DE practically from the outset and galaxies would not have formed.
By inspecting the left hand side of Eqs. (\ref{diff1}) and
(\ref{diff2}), it must be a function of the energy densities
multiplied by a quantity with units of inverse of time for which
we take the Hubble factor as it seems a natural choice. Therefore,
we end up with an expression such as $Q=Q(H\rho_{m},H\rho_{D})$.
If we expand this function as a power law and keep just the first
term, we have $Q\simeq \lambda_{m}H\rho_{m}+\lambda_{D}H\rho_{D}$.
Given the absence of information about the coupling, it makes
sense to work with just one parameter, so the three possible
choices are: $\lambda_{m}=0$, $\lambda_{D}=0$ and
$\lambda_{m}=\lambda_{D}$. Here in this paper we consider the
latter choice form \cite{Wang:2007ak}
\begin{equation}\label{Q}
Q=3b^{2}H(\rho_{m}+\rho_{D}),
\end{equation} where $b^{2}$ is the coupling constant and $3H$ is attached for dimensional consistency.
With this choice for $Q$, Eq. (\ref{rpunto}) takes the form
\begin{equation}\label{r}
\dot{r}=3Hrw_{D}+3b^{2}H(1+r)^{2}.
\end{equation}

The positive term $b^{2}$ gauges the energy transition from DE to DM. A negative $b^{2}$ would violate the second law of thermodynamics. $b^{2}$ is usually taken within the range $[0, 1]$
\cite{zhang}. $b^{2}=0$ represents the absence of interaction whereas $b^2=1$ implies a complete transfer of
energy from DE to DM.  Using the latest observations (golden SN Ia, the shift parameter of CMB and the BAO) and combining them
with the lookback time data we have that $b^2$ could be as large as $0.2$ (see \cite{Feng:2008fx}) although a value of $b^2< 0.04$ is favoured.

Now, inserting Eqs. (\ref{dotrho1}) and (\ref{Q}) into Eq. (\ref{diff2})
and make use of Eq. (\ref{u1}) we get the equation of state parameter for the
interacting GDE
\begin{equation}\label{wD2}
w_D=-\frac{1}{2-\Omega_D}\left(1+\frac{2b^2}{\Omega_D}\right).
\end{equation}
For late time, when $\Omega_D\rightarrow 1$, we see that $w_D$ crosses the phantom
divide line because $w_D=-(1+2b^2)<-1$ irrespective of the value of
coupling $b^2$. When we consider the present epoch, this is, when $\Omega_{D,0}=0.72$, the phantom crossing will take place as long as $b^2>0.1$. This is compatible with current observations (\cite{Feng:2008fx}).

The differential equation that governs the dynamics of the interacting GDE was obtained in \cite{Sheykhi:2011nb}
\begin{equation}\label{Omegaprime3} \frac{d\Omega_D}{dz}=-\frac{3}{2(1+z)}\Omega_D\left[1-\frac{ \Omega_D}{2-
\Omega_D}\left(1+\frac{2b^2}{\Omega_D}\right)\right].
\end{equation}
The evolution of $w_D$ and $\Omega_{D}$ is depicted in Figs. 7 and 8 for different values of $b^2$. We can see in Fig. 7 the phantom crossing for $b^{2}>0.1$

The deceleration parameter for the interacting GDE is given by
\begin{equation}\label{q2}
q=\frac{1}{2}+\frac{3}{2}w_{D}\Omega_{D}=\frac{1}{2}-\frac{3}{2}\frac{\Omega_{D}}{(2-\Omega_D)}\left(1+\frac{2b^{2}}{\Omega_{D}}\right)
\end{equation} and its variation with respect to $z$ and the effect of the interaction in the cosmic evolution for different values of the coupling is shown in Fig. 9. We can see that the cosmic acceleration starts earlier when the interaction is taken into account as DE dominates earlier. Furthermore, the larger the coupling, the earlier the acceleration of the Universe starts. However, the cases with smaller coupling will get larger
acceleration finally in the far future.

Exactly as we did in the non-interacting case, we now proceed to establish the correspondence between the interacting GDE and the kinetic k-essence scalar field, i.e. we identify their respective energy densities and equate $w_{D}$ in Eq. (\ref{wD2}) with
$w_\phi$.

Although the GDE model allows the phantom crossing for values of $b^2>0.1$, we cannot use the kinetic k-essence to mimic this behaviour. The reason behind this is that it is impossible for a purely kinetic k-essence to achieve the phantom crossing as shown in \cite{Sen:2005ra}. Moreover, one cannot obtain k-essence phantom models without quantum instabilities \cite{Abramo:2005be}. Therefore, we shall reconstruct $F(X/X_{0})$ for $b^2<0.1$. This is not too restrictive because in spite of the fact that values of $b^2>0.1$ are marginally allowed \cite{Feng:2008fx}, most observations give a value for the coupling parameter of $b^2<0.04$ \cite{Feng:2008fx} or even a smaller $b^2<0.025$ \cite{ich}.

In order to obtain $F(z)$, we shall use Eq. (\ref{Fdez}), where $\omega_{D}$ and $\Omega_{D}$ are respectively given by Eqs. (\ref{wD2}) and (\ref{Omegaprime3}), and $H$ is expressed as \cite{Hu:2006ar}
\begin{figure}[htp]
\psfrag{z}{$z$} \psfrag{b2}{$b^{2}$} \psfrag{F}{$F$}
\includegraphics[scale=0.92]{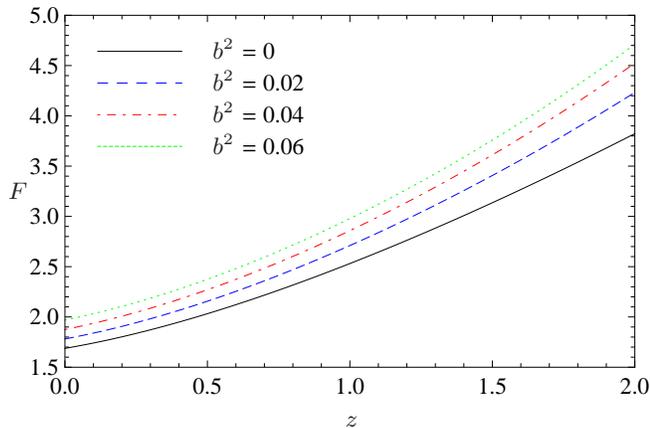}
\caption{Variation of $F(z)$ in units of $\frac{\alpha^{2}}{3M_{P}^{2}}$ for the interacting ghost dark energy model.}
\end{figure}

\begin{figure}[htp]
\psfrag{z}{$z$} \psfrag{b2}{$b^{2}$} \psfrag{X0}{$X_{0}$} \psfrag{X}{$X$}
\includegraphics[scale=0.87]{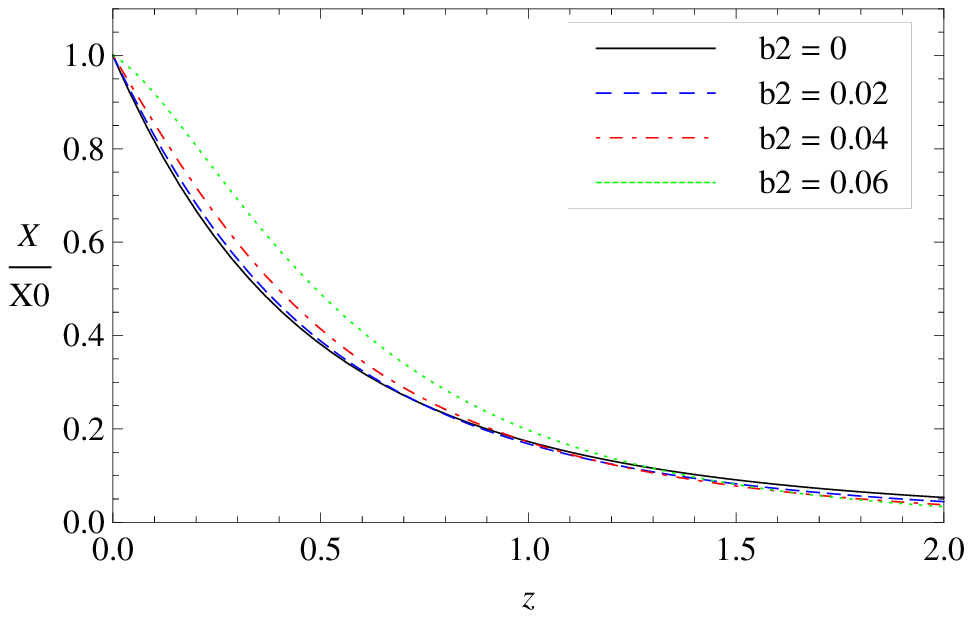}
\caption{Variation of $\frac{X}{X_{0}}(z)$ in units of $M_{P}^{2}H_{0}^{2}$ for the interacting ghost dark energy model.}
\end{figure}
\begin{equation}\label{Hint}
H=H_{0}e^{3/2\int_{0}^{x}\left(  1+{\frac{\omega_{D}}{1+r}}\right)
dz}
\end{equation} which is different from Eq. (\ref{Fridmann1}) because the interaction modifies the expansion history of the
Universe.
On the other hand, combining Eqs. (\ref{Omega}), (\ref{GDE}),(\ref{wD2}), and (\ref{Hint}) yields $X$ as a function of $z$
\begin{equation}\label{Xdezint}
\frac{X}{X_{0}}(z)=\left\{\frac{H^{2}\left[2b^{2}+\Omega_{D}(\Omega_{D}-1)\right]}{(2-\Omega_{D})(1+z)^{3}\left[\Omega_{m,0}+\frac{2(\Omega_{D,0}+b^{2}-1)}{(2-\Omega_{D,0})}\right]}\right\}^{2}.
\end{equation}
We are now in a position to obtain the function $F=F(X/X_{0})$ numerically.

Selected curves for $F(z)$ and $X/X_{0}(z)$ corresponding to different values of the coupling are shown in Figs. 10 and 11, respectively. The reconstructed $F(X/X_{0})$ is plotted in Fig. 10. We learn from these figures that the reconstructed scalar field has the same
dynamics as the non-interacting case, which is due to the smallness of the coupling that gauges the interaction in the GDE model. This was expected because otherwise the GDE model would deviate significantly from the
concordance model, making it incompatible with observations
\cite{matter perturbations}.

\begin{figure}[htp]
\psfrag{z}{$z$} \psfrag{b2}{$b^{2}$} \psfrag{X0}{$X_{0}$} \psfrag{X}{$X$} \psfrag{F}{$F$}
\includegraphics[scale=0.95]{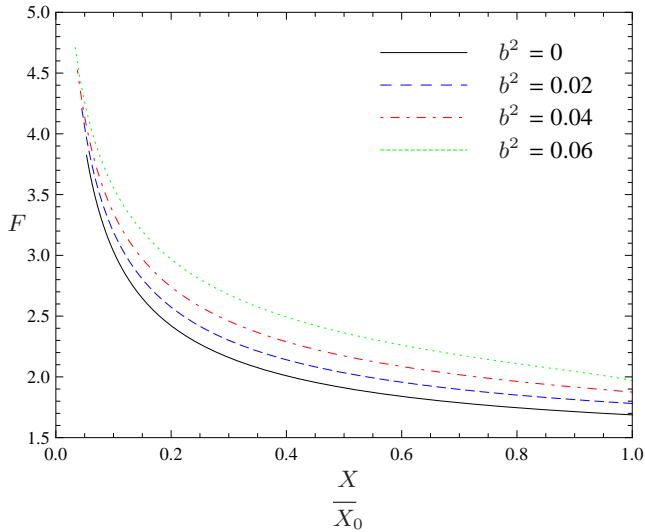}
\caption{Reconstructed $F(X/X_{0})$ for the interacting ghost dark energy model.}
\end{figure}
\begin{figure}[htp]
\psfrag{z}{$z$}
\psfrag{rhomb2}{$\Omega_{m}$, $b^2$}
\psfrag{rhoDb2}{$\Omega_{D}$, $b^2$}
\includegraphics[scale=0.955]{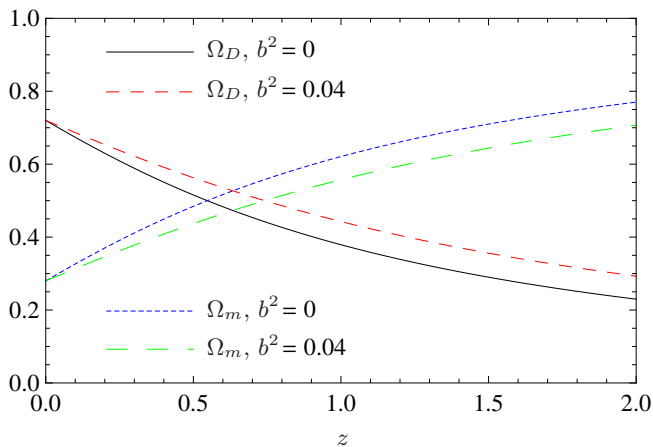}
\caption{Variation of $\Omega_{D}$ and $\Omega_{m}$ with respect to $z$ for the interacting ghost dark energy model.}
\end{figure}

\begin{figure}[htp]
\psfrag{z}{$z$}
\psfrag{rhomb2}{$\rho_{m}$, $b^2$}
\psfrag{rhoDb2}{$\rho_{D}$, $b^2$}
\includegraphics[scale=0.985]{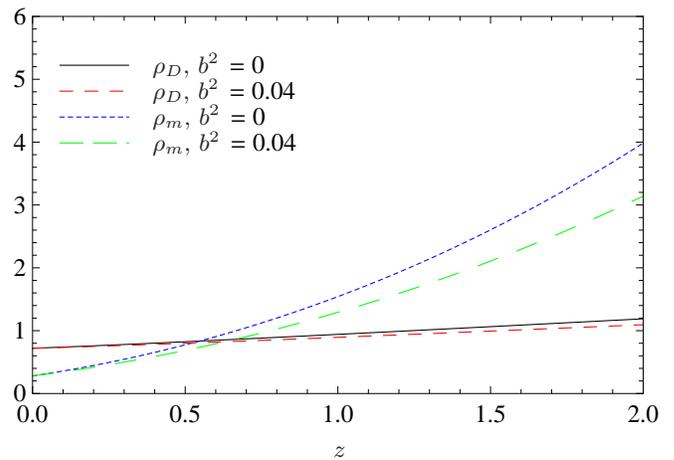}
\caption{Variation of $\rho_{D}$ and $\rho_{m}$ in units
of $\rho_{cr,0}$ with respect to $z$ for the interacting ghost dark energy model.}
\end{figure}

\begin{figure}[htp]
\psfrag{z}{$z$} \psfrag{r}{$r$}
 \psfrag{b2}{$b^{2}$}
\includegraphics[scale=0.97]{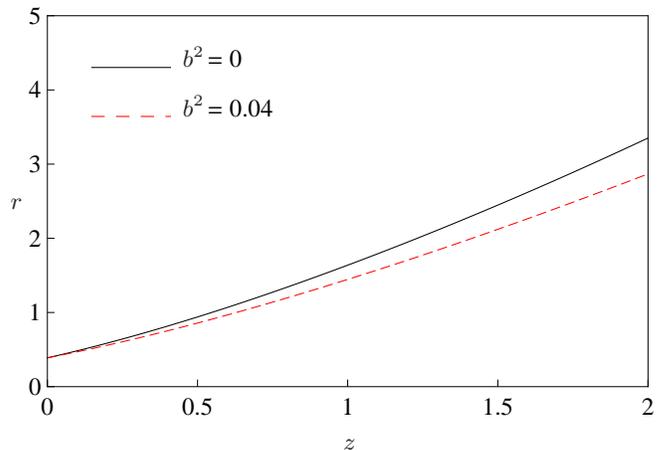}
\caption{Variation of the ratio r with respect to $z$ for the interacting ghost dark energy model.}
\end{figure}
In Fig. 11 the effect of the interaction between GDE and DM is displayed and we see that $\Omega_{D}$ evolves at a faster rate in the interacting case. Moreover, Fig. 14 shows the
point where $\rho_{D}$ and $\rho_{m}$ cross, $\rho_{D}=\rho_m$, takes place
earlier when the interaction is present. This latter feature is
appreciated in more detail in Fig. 15 where the dependence of the
ratio $r \equiv \rho_m/\rho_{D}$ with respect to $z$ is
depicted.  This ratio $r$ decreases monotonously with
the expansion and varies slowly at the present epoch. This reduction is
slower when the interaction is considered, implying that in
this scenario the coincidence problem gets substantially
alleviated and besides, that DE is decaying into DM in recent
epochs.

\section{Conclusions}\label{con}

The Veneziano ghost dark energy (GDE) has been recently proposed as a new model of dark energy. The crucial ingredient, the Veneziano ghost, shows important non-trivial physical effects in a dynamical background such as our FRW expanding Universe, or in a spacetime with a non-trivial topology. These small effects produce a vacuum energy density $\rho_{D} \sim \Lambda_{QCD}^{3}H \sim (10^{-3} {\text eV})^{4}$, where $\Lambda_{QCD}$ is the QCD mass scale and $H$ is the Hubble parameter. Given that $\Lambda_{QCD}\sim100MeV$ and $H\sim10^{-33}eV$, this $\rho_{D}$ has the right value to drive the current acceleration of the Universe. The advantages of the GDE with respect to other dark energy models include the absence of the fine tuning problem and the fact that it can be completely explained within the standard model and general relativity, without recoursing to any new field, new degree(s) of freedom, new symmetries  or modifications of general relativity.

For some time, to consider scalar field models of dark energy as
effective theories of an underlying theory of dark energy has been a mainstream idea. In this spirit, we have investigated the prospects of reproducing the behaviour of the GDE model with the kinetic k-essence scalar field model with an appropriate Lagrangian, also studying the case of a possible interaction between dark matter and dark energy. For this purpose, we have established a correspondence between the GDE and the kinetic k-essence field, by identifying their respective energy densities and equations of state, and reconstructed the kinetic k-essence GDE model in the region $w>-1$. Describing the GDE model in a scalar field framework provides a more fundamental representation of the dark component. Finally, we have executed an exhaustive
analysis of its evolution and dynamics where we have found that it is possible to obtain an equivalent description of the dark fluid in the case of a purely kinetic k-essence Lagrangian and the model is able to reproduce the correct behaviour of the cosmological background for $z\leq2$. In addition, the coincidence problem gets substantially alleviated when the interaction between dark energy and dark matter is taken into account.

\begin{acknowledgments}
This work was supported by MICINN (Spain) under research
project no. FIS2008-06332 and by Fundaci\'on Ram\'on Areces.
\end{acknowledgments}

\end{document}